\documentstyle[12pt,manuscript,aps,epsf,colordvi]{revtex}  
\begin{document}
\draft
\title{$d^{\prime }$ production in heavy ion collisions}
\author{S.M. Kiselev, M.I. Krivoruchenko and B.V. 
Martemyanov}
\address{Institute for Theoretical and Experimental 
Physics, B.Cheremushkinskaya 25\\ 
117259 Moscow, Russia} 
\author{Amand Faessler and C. Fuchs}
\address{ Institut f\"ur Theoretische Physik, Universit\"at 
T\"ubingen, Auf der Morgenstelle 14 \\
D-72076 T\"ubingen, Germany}
\maketitle
\begin{abstract}
The production of $d^{\prime }$ dibaryons in heavy ion 
collisions due to the elementary 
process $NN\rightarrow d^{\prime }\pi $ is considered. 
The $NN\rightarrow d^{\prime }\pi $ 
cross section is estimated using the vacuum $d^{\prime }$ 
width $\Gamma _{d^{\prime }}\approx 0.5$ MeV 
extracted from data on the double 
charge exchange reactions on nuclei.  
The $d^{\prime }$ production rate per single collision
of heavy ions is estimated at an incident beam energy of 
1 $A\cdot$GeV within the framework of the Quantum Molecular 
Dynamics transport model. We suggest to 
analyse the invariant mass spectrum of the $NN\pi $ 
system in order to search for an abundance of 
events with the invariant mass of the $d^{\prime }$ 
dibaryon. The $d^{\prime }$ peak is found to exceed 
the statistical fluctuations of the background at a 
$6\sigma$ level for $2\;10^{5} \cdot A$ central 
collisions of heavy ions with the atomic number $A$. 
\vspace{0.5cm}\\{\bf keywords}: heavy ion collisions, 
dibaryons
\end{abstract}

\section{Introduction}

In the last years the possibility of the existence and the properties of 
a narrow dibaryon $d^{\prime}\; 
(T=0,J^p=0^{-})$ have been widely discussed both from 
the experimental  
and theoretical points of views \cite{bi1} - \cite{bro}. So 
far, the narrow resonance structure is 
unambiguously established only in the double charge 
exchange (DCE) reactions on nuclei from 
$^7Li$ to $^{56}Fe$ \cite{w}. The peaks in the forward 
DCE reaction cross sections at the 
incident kinetic energy of pions $T_\pi \approx 50$ MeV 
are well described assuming that the 
dominant contribution of this process stems from a $d^\prime$ 
dibaryon with a mass $M_{d^\prime} = 
2063$ MeV and a free decay width $\Gamma_{d^\prime} 
\approx 0.5$ MeV. The $d^{\prime}$ 
mechanism is presently the only one which allows to explain the 
observed narrow structures. It 
is therefore important to find an evidence for a $d^{\prime}$ 
dibaryon in other reactions. 
At CELSIUS (Uppsala), the reaction 
$pp \rightarrow pp\pi^+\pi^-$ was studied \cite{bro}. The analysis of the 
$pp\pi^-$ invariant mass spectrum revealed a $4\sigma$ 
peak at $M \approx M_{d^\prime}$. 
The reported data are, however, still preliminary. Further 
plans to search for a $d^\prime$ 
dibaryon include proton-proton collisions and photo- and 
electro-production on deuterons.

The existence of narrow dibaryon resonances has 
important consequences for physics of 
dense nuclear matter. When the density is increased beyond a critical 
value, the formation of a Bose 
condensate of the dibaryons becomes energetically 
favourable. This phenomenon has been studied in 
Refs. \cite {FBKM} - \cite{BFK}.

In the present work, the possibility of the $d^{\prime }$ production 
in heavy ion collisions is investigated for incident 
beam energies around 1 $A\cdot$GeV. 
Such a beam is, {\it e.g.}, available at SIS at GSI (Darmstadt). 
We investigate the $d^{\prime }$ production 
rate and the $pp\pi^{-}$ background within the framework 
of the Quantum Molecular 
Dynamics (QMD) transport approach \cite{aich91} 
which is well established and describes successfully 
many observables of heavy ion reactions 
\cite{Ai86} - \cite{kis}. 

In the next Sect., the $NN\rightarrow d^{\prime }\pi $ 
cross section is calculated
using as an input value the vacuum $d^{\prime }$ width 
$\Gamma _{d^{\prime
}} \approx 0.5$ MeV extracted from the DCE reactions. We expect 
$2\;10^{-4}\;d^{\prime }$ per single $NN$ collision. 
In Sect. 3 we start with a qualitative discussion of the 
$d^{\prime}$ production dynamics in heavy ion collisions. Using 
reasonable assumptions, we give an estimate for 
the $d^{\prime}$ production rate and for the 
number of heavy ion collisions needed to
separate the $d^{\prime}$ peak from the background.

In Sect. 4, the $d^{\prime }$ production  
in heavy ion collisions is studied with the 
use of the QMD simulations. The 
$d^{\prime }$ production rate is calculated for 
central collisions of the ions $^{20}Ne$,
$^{59}Ni$, $^{108}Ag$, and $^{197}Au$.
The invariant mass distribution of the background $pp\pi^{-}$ events 
is analysed in Sect. 5. 
We found that it is possible to distinguish a $6\sigma$ $d^{\prime 
}$ peak from the analysis
of at least $2\;10^5 \cdot A$ collisions. Corresponding experiments at 
the SIS  have collected up to now 
about $10^8$ events. Thus the possibility for the search for the 
$d^{\prime }$ dibaryon in heavy ion collisions is 
apparently a realistic aim.

\section{Cross section for $d^{\prime }$ production in 
nucleon-nucleon collisions}

The possible $NNd^{\prime }\pi $ couplings are carefully 
analysed in Ref. 
\cite{szc}. Thus we start with the form of the $NNd^{\prime 
}\pi $ vertex
suggested in this work 
\begin{equation}
T_{NNd^{\prime }\pi }=g\psi _1^TC\gamma _5\tau _i\psi 
_2\pi_id^{\prime }.
\label{vertex}
\end{equation}
Here, $C=\gamma _2\gamma _0$ and $g$ is a form factor 
which depends on the momenta
of the particles involved.
The two nucleons in the vertex (\ref{vertex}) are in the 
relative $^1S_0$ state,
so they are affected by the long-distance attraction resulting 
in a virtual pole and by the short-distance repulsion. The final-state 
interaction of two nucleons substantially modifies the 
$d^{\prime }$ decay
rate. On the other hand, the initial-state interaction 
of the nucleons in the reaction $%
NN\rightarrow d^{\prime }\pi $ modifies the cross 
section.

First we assume a constant form factor $g$ and neglect the 
interaction of the nucleons. Then we discuss in details the 
final- and
initial-state interactions and the influence of the nontrivial 
momentum dependence of the form factor $g$.

The total width $\Gamma _{d^{\prime }}^{(0)}$ of the 
$d^{\prime }
$ dibaryon for a constant $g$ with no final-state 
interactions can be calculated as  
\begin{equation}
\label{gamma0}\Gamma _{d^{\prime }}^{(0)}=g^2\frac 
3{64\pi
^2}\sqrt{\mu M_{d^{\prime }}}T_{\pi ,\max }^2.
\end{equation}
Here, $T_{\pi ,\max }=((M _{d^{\prime }}-\mu )^2-
4m^2)/(2M _{d^{\prime }})\approx 
45$ MeV is the maximum
pion kinetic energy in the $d^{\prime }$ decay, $\mu $, 
$m$, and $M_{d^{\prime }}$ are the
pion, nucleon, and dibaryon masses, respectively. Due to 
the isotopic invariance, the partial decay widths are 
given by
$$\Gamma _{d^{\prime } \rightarrow pp \pi^-}^{(0)}=
\Gamma _{d^{\prime } \rightarrow pn \pi^0}^{(0)}=
\Gamma _{d^{\prime } \rightarrow nn \pi^+}^{(0)}=
\frac {1}{3}\Gamma _{d^{\prime }}^{(0)}.
$$
The same relations hold true for the widths corrected to the 
final-state interactions.

The total cross section of the dibaryon production $\sigma 
_{pp\rightarrow
d^{\prime }\pi^+ }^{(0)}$ equals 
\begin{equation}
\label{sigma0}\sigma _{pp\rightarrow d^{\prime }\pi 
^+}^{(0)}=g^2\frac 1{8\pi
}\frac kp
\end{equation}
where $k$ and $p$ are the pion and nucleon momenta in 
the center-of-mass
frame. In the different isotopic channels, 
the $d^{\prime }$ production cross sections 
are related by
$$
\sigma _{pp\rightarrow d^{\prime }\pi ^+}^{(0)} =
\sigma _{nn\rightarrow d^{\prime }\pi ^-}^{(0)} =
2 \sigma _{pn\rightarrow d^{\prime }\pi ^0}^{(0)}.
$$
The same relations are valid for the cross sections 
corrected by the initial-state interactions.
In terms of the $d^{\prime }$ width, the $pp\rightarrow 
d^{\prime }\pi ^+$ 
cross section takes the form 
\begin{equation}
\label{siggam0}\sigma _{pp\rightarrow d^{\prime }\pi^+ 
}^{(0)}=\Gamma
_{d^{\prime }}^{(0)}\frac{8\pi }3\frac 1{\sqrt{\mu 
M_{d^{\prime }}}
}\frac 1{T_{\pi ,\max }^2}\frac kp.
\end{equation}

If the final-state interaction of the 
nucleons in the $d^{\prime }$ decay and the
initial-state interaction of the nucleons in the $d^{\prime }$ 
production are taken into account, we obtain both, 
a corrected width $\Gamma_{d^{\prime}}$ 
and a corrected cross section $\sigma_{pp\rightarrow
d^{\prime }\pi^+ }$:
\begin{equation}
\label{corrections}\Gamma _{d^{\prime }}=\Gamma
_{d^{\prime }}^{(0)}\eta _F\;\;\;\;{\rm and}\;\;\;\;\sigma
_{pp\rightarrow d^{\prime }\pi^+ }=\sigma 
_{pp\rightarrow d^{\prime }\pi^+
}^{(0)}\eta _I. 
\end{equation}
The corrected cross section contains, as compared to 
Eq. (\ref{siggam0}), an
additional factor $\eta _I/\eta _F$ 
\begin{equation}
\label{siggam}\sigma _{pp\rightarrow d^{\prime }\pi^+ 
}=\Gamma _{d^{\prime
}}\frac{8\pi }3\frac 1{\sqrt{\mu M_{d^{\prime
}}}}\frac 1{T_{\pi ,\max
}^2}\frac kp\frac{\eta _I}{\eta _F}. 
\end{equation}

In contrast to Ref. \cite{szc} where the factors $\eta _I$ and $\eta _F$ 
have been determined form different sources, here we use a method 
which allows to calculate both, $\eta _I$ and $\eta _F$, consistently. 
The common scale of $\eta _I$ and 
$\eta_F $ remains in 
our method unfixed due to the unknown asymptotic behaviour 
of the nucleon scattering phase shifts at high energies. The 
common scale factor, however, drops out from the ratio $\eta _I/\eta _F$
entering into Eq. (\ref{siggam}). The present method is based on the 
integral
representation of the Jost function \cite{gw} 
\begin{equation}
\label{jost}D_I(T)=\exp \left( -{\frac 1\pi \int_0^\infty 
\frac{\delta
(T^{\prime })}{T^{\prime }-T}dT^{\prime }}\right).
\end{equation}
Here, $T$ is the nucleon kinetic energy in the laboratory 
frame.
The value of $T$ is connected to the invariant energy 
$s=(p_1+p_2)^2$ by the relation $s=4m^2+2mT.$
In Eq. (\ref{jost}), $\delta (T)$ is the phase shift in the 
dominant $^1S_0$
partial wave state of the two nucleons in the $d^{\prime}$ 
decay.

It is known from scattering theory that the 
influence of the final-state interaction on the
outgoing particles can be taken into account by dividing the 
bare
amplitude by the Jost function. The same prescription is 
valid for the
initial-state interaction. The amplitudes with two external 
nucleon lines get
therefore factors $1/D_I(T).$

The enhancement factor $\eta _I$ entering 
Eq. (\ref{corrections}) is simply 
$$
\eta (T)=\frac 1{|D_I(T)|^2},
$$
while the factor $\eta _F$ can be calculated by averaging 
the same function $%
\eta (T)$ over the phase space distribution of two nucleons 
originating from the decay $d^{\prime }\rightarrow NN\pi 
$.

The phase $\delta (T)$ is experimentally known up to 
$T=1000$ MeV \cite{arn}
and can be extrapolated to infinity in a model dependent 
way. We use the
experimental data in the interval $T=0\div 1000$ MeV and 
results of the
model \cite{tabak} for $T>1000$ MeV. The uncertainties 
in the asymptotic behaviour of the phase
do, however, not significantly influence 
the ratio $\eta _I/\eta _F$. The 
function $\beta (T)=\frac 12\ln
\left( \eta (T)/\eta (0) \right) $ is shown in Fig. 1 for the 
interval $T=0 \div 1000$ MeV (solid curve). 
It is seen that it rapidly falls off from $\beta (0)=0$ to a minimum
value of $\beta (T_1)\approx -3.3$ at a kinetic energy of $T_1\approx 600$ 
MeV. In the interval $%
T=700\div 1000$ MeV, $\beta (T)$ is a slowly varying 
function. For comparison, we plot in Fig. 1 the function $\beta(T)$
calculated in the effective-range approximation (see Appendix). The ratio 
$\eta _I/\eta _F$ in our approach becomes

\begin{equation}
\label{etta}
\frac{\eta _I}{\eta _F}\approx \frac 15. 
\end{equation}
Inserting  $M_{d^{\prime}}=2063$ MeV and $\Gamma 
_{d^{\prime}}=0.5$ MeV, 
as suggested by the interpretation of the DCE
reactions \cite{w}, into Eq. (\ref{siggam}) we obtain for the 
cross section
\begin{equation}
\label{sigma}\sigma _{pp\rightarrow d^{\prime }\pi^+ 
}=300\frac kp \;\mu {\normalsize b}. 
\end{equation}

In the reaction $NN\rightarrow d^{\prime }\pi $, the 
threshold energy is
$T_0=710$ MeV. At $T\approx 730$ MeV, and 
$k/p\approx 1/10$ 
one obtains $\sigma_{pp\rightarrow d^{\prime }\pi^+ 
}=30 \; \mu$b.
The cross section (\ref{sigma}) should be compared to the
experimental data on the $d^{\prime }$ production 
in the $pp \rightarrow pp\pi^+\pi^-$ reaction \cite{bro}. At
$T = 750$ MeV, where the total cross section of the
$pp \rightarrow pp\pi^+\pi^-$ reaction is about $10\div 20 \; 
\mu$b, 
the $d^{\prime }$ contribution is estimated to be at a 
$7\%$ level.
It means that at this energy
$\sigma _{pp\rightarrow d^{\prime }\pi^+ } \approx 
15\cdot
0.07\cdot 3 \approx 3 \;\mu$b, {\it  i.e.} the empirical value 
is about ten times smaller 
than the value given by Eq. (\ref{sigma}). Hence, we are
forced to conclude that the constant form factor $g$ does 
not allow to calculate the cross section 
$\sigma _{pp\rightarrow d^{\prime }\pi^+ } $ correctly from the 
width $\Gamma _{d^{\prime }}$.
Thus, we need to introduce a form factor to suppress
the $NNd^{\prime }\pi $ vertex at high $NN$ energies.
One  possible choice of the form factor is the
following
\begin{equation}
g^2 = \frac{g^{\prime 2}}{1 + \frac{T^2}{m_{FF}^2}}
\label{ff}
\end{equation}
where $m_{FF}$ is a cut-off mass.
The required suppression allows to determine the mass 
parameter
 $m_{FF}\approx 240$ MeV.
Finally, we get for the cross section 
\begin{equation}
\label{sigmaff}\sigma _{pp\rightarrow d^{\prime }\pi 
^+}=
\frac{300}{(1 + \frac{T^2}{m_{FF}^2})}\frac kp \;\mu 
{\normalsize b}. 
\end{equation}
Fig. 2 shows the cross section versus the nucleon kinetic energy 
from the $d^{\prime }\pi$ threshold up to $T=1000$ MeV.
 
It is seen from Eq. (7) that the enhancement factor 
$\eta(T)$ approaches unity with 
increasing T. In the potential framework, the final- and 
initial-state interaction effects disappear 
with increasing the energy. In our case ${\eta _I}/{\eta 
_F}\approx  1/5$ and $\eta_I$ is almost 
constant in the interval $T=700\div 1000$ MeV. One can 
thus assume that at $T=700\div 
1000$ MeV initial-state interaction effects are essentially 
switched off. 
The result ${\eta _I}/{\eta _F}\approx  1/5$ should be 
compared to the result of Ref. 
\cite{szc}: $\eta_F \approx 5$, $\eta_I \approx 1/10$, and 
$\eta_I/\eta _F \approx 1/50$. 
The order-of-magnitude difference is apparently related to 
the effect of a $d^{\prime}$ form 
factor implicitly included in calculations of the enhancement factor 
$\eta_I$ performed in Ref. \cite{szc}. The effective-range approximation
gives $\eta_F \approx 3$ and $\eta_I \approx 1$ (see Appendix).

\section{Naive estimate for $d^{\prime }$ production in 
heavy ion collisions}

In the previous section the cross section 
$pp\rightarrow d^{\prime }\pi^+ $ has been estimated 
using thereby as an 
input value the vacuum $d^{\prime }$ width $\Gamma 
_{d^{\prime}}\approx 0.5$ MeV. This 
cross section turned out to be rather small. 
For a kinetic energy of $T = 1$ GeV of 
the first nucleon in the rest frame of 
the second nucleon the $d^{\prime}$ production cross 
section reaches a value of $\sigma _{pp\rightarrow d^{\prime 
}\pi^+ } \approx 10 \;\mu$b (see Fig. 2). 
The total nucleon-nucleon cross section at the same energy 
is about $\sigma_{NN}^{tot} \approx 40$ 
mb. The ratio between the total $d^{\prime }$ production 
cross section averaged over the 
isospin states of the initial nucleons and the total 
nucleon-nucleon cross section is about
\begin{equation}
\label{ksi}\xi =\frac{\frac{3}{4}\sigma _{pp\rightarrow 
d^{\prime }\pi^+ }}{\sigma
_{NN}^{tot}}\approx 2 \; 10^{-4}.
\end{equation}
We thus expect $\xi \approx 2 \; 10^{-4}$ $d^{\prime 
}$ dibaryons to be produced per single $NN$ 
collision.

In heavy ion collisions the number of the elementary 
$NN$ scatterings is already at intermediate energies relatively 
large and thus the 
$d^{\prime }$ production rate is 
enhanced. The mean free path $\ell _f\approx 1$ fm of a
nucleon in a heavy nucleus is small as compared to the nuclear size. 
Therefore a nucleon from the colliding 
nucleus undergoes several scatterings with nucleons from the 
target nucleus, loses its energy and 
finally cannot produce a $d^{\prime }$ dibaryon ($\sqrt s < 
M_{d^{\prime }}$) anymore. A naive 
estimate of the number of the $d^{\prime }$ dibaryons per a single 
$AA$ collision yields
\begin{equation}
\label{OOO}
n_{d^{\prime }}=\xi A \approx 0.04 \left( \frac \xi {2 \; 
10^{-4}}\right)
\left( \frac A{200}\right). 
\end{equation}
The $d^{\prime }$ production rate in symmetric and central 
heavy ion collisions is roughly enhanced by a factor $A$ 
corresponding to the number of the projectile/target nucleons 
as compared to $pp$ collisions. 
The background in heavy ion collisions is more complicated. 

We wish to select $NN\pi$ events to test the invariant mass 
distribution with respect to the 
presence of a peak caused by the $d^{\prime }$  dibaryon. 
For total number of the pions 
produced per single $AA$ collision at an energy of 1 
$A\cdot$GeV, we assume $N_\pi \approx A/4.$ 
This number is an upper limit for the pion production rate
in the central heavy ion collisions, 
which coincides roughly with the prediction 
of the so-called Harris systematics \cite{harris}
at that energy. 
However, recent experiments \cite{pelte97} indicate that 
pion production is suppressed in 
heavy systems with respect to the Harris systematics. The 
total number $N_t$ of those $NN\pi$ events 
which have to be analysed is then
\begin{equation}
\label{notriplet}N_t=2A^2N_\pi N_C \approx 4\; 
10^6\left( \frac
A{200}\right) ^3 N_C,
\end{equation}
with $N_C$ being the number of nucleus-nucleus collisions.

For the further discussion we denote 
with $n_F({\bf p})$ and $n_B({\bf k})$ the momentum space distribution 
functions of the outgoing nucleons and pions, respectively. 
The number density of $NN\pi$ triplets with respect to the $NN\pi$ 
invariant mass $M_t$ is given by
\begin{equation}
\label{numberdensity}\frac{1}{N_C} 
\frac{dN_t}{dM_t^2}=\frac{g_{nucl}^2g_{pion}}{2!}
\int \frac{d {\bf p}_1}{(2\pi)^3}\frac{d {\bf p}_2}{(2\pi)^3}
\frac{d {\bf k}}{(2\pi)^3}
\delta \left((p_1 + p_2 + k)^2 - M_t^2\right) \, n_F({\bf p}_1)\, n_F({\bf 
p}_2)\, n_B({\bf k})
\end{equation}
where $g_{nucl} = 2\times2$, $g_{pion} = 3$ are 
statistical factors for nucleons and pions. 
Since we are interested in the values of $M_t$ close to the 
$NN\pi$ threshold, one can write
\begin{equation}
\label{numberdensity1}\frac{1}{N_C} 
\frac{dN_t}{dM_t}\;\approx\;
\frac{g_{nucl}^2g_{pion}}{2!}\frac{1}{(
2\pi)^6}
\Phi_3(M_t)
\int \frac{d {\bf P}}{(2\pi)^3}\frac{M_t}{\sqrt{M_t^2 + 
{\bf P}^2}}\, 
n^{2}_{F}({\frac{m}{M_t}\bf P}) \, n_B({\frac{\mu}{M_t}\bf 
P})
\end{equation}
where
\begin{equation}
\label{phase2} \Phi_3(M_t) = \int d {\bf p}_1 d {\bf p}_2 
d {\bf k} \,\delta^4 (p_1 + p_2 + k - P)
\approx\frac{\pi^3}{2}({\mu}M_t)^{3/2}\,
T^{2}_{\pi ,\max }(M_t)
\end{equation}
is essentially the phase space of the $NN\pi$ system. Here, 
$P=(\sqrt{M_t^2+{\bf P}^2},{\bf 
P})$, $m$ and $\mu$ are nucleon and pion masses, and 
$T_{\pi ,\max }=((M_t - \mu )^2 - 4m^2)/(2M_t)$
is the maximum kinetic energy of the pions in the 
center-of-mass frame of the $NN\pi$ 
system for a given invariant mass $M_t$.

Near the $NN\pi$ threshold, the dependence of the density 
$dN_t/dM_t$ on the invariant mass 
$M_t$ is mainly governed by the factor $T^{2}_{\pi ,\max }(M_t)$. 
One can therefore write
\begin{equation}
\label{phalabel}\frac{1}{N_t}\frac{dN_t}{dM_t}\;\approx
\;
\frac{T^{2}_{\pi ,\max }(M_{d'})}{\int_{2{\;}\rm 
GeV}^{2.3{\;}\rm GeV}{T^{2}_{\pi ,\max}(M_t) dM_t}}
\;\approx\; 4\;10^{-4}\;{\rm MeV}^{-1}.
\end{equation}
The density of the background events in vicinity of the 
$d^{\prime}$ peak is estimated as 
\begin{equation}
\label{phasespace}\frac{1}{N_C} 
\frac{dN_t}{dM_t}{\;\approx \;}
\frac{2A^2N_{\pi}}{N_t}\frac{dN_t}{dM_t}{\;\approx 
\;}
\frac{1600}{\rm MeV}\left( \frac A{200}\right) ^3.
\end{equation}
In the case of the $^{197}Au\;+\;^{197}Au$ collisions, 
we thus obtain about $1600$ background events per $1$ MeV bin
of spectrum of the invariant mass $M_t$.

The actual bin width $\Delta M_t$ which we are interested in 
is given by the maximum of the $d^{\prime }$ width 
$\Gamma _{d^{\prime }}\approx 0.5$ MeV and the 
experimental resolution $(\Delta 
M_t)^{\exp }$. The number of the background events in 
this interval, 
\begin{equation}
\label{III}
\frac{\Delta 
N_t}{N_C}=\frac{2A^2N_{\pi}}{N_t}\frac{dN_t}{dM_t
}\Delta M_t
{\;\approx \;} 1600\left( \frac A{200}\right) ^3
\left( \frac{\Delta M_t}{1\;{\rm MeV}}\right),
\end{equation}
should be compared to the number of the $d^{\prime }$ 
events given by Eq. (13).

Suppose we have $N_C$ collisions and we want to have a 
statistically significant
peak at a $n_\sigma \sigma $ level with $n_\sigma \geq 6$. 
We should require
\begin{equation}
\label{PPP}
n_\sigma \sqrt{N_C\Delta n_t}\leq N_Cn_{d^{\prime }}, 
\end{equation}
with $\Delta n_t= \Delta N_t/N_C$ being the number of the 
$NN\pi$ events 
per single $AA$ collision in the interval $\Delta M_t$. We get
\begin{equation}
\label{AAA}
N_C\geq n_\sigma ^2\frac{\Delta n_t}{n_{d'}^2} 
{\;\approx\;} 4\; 10^7
\left( \frac{n_\sigma }{6}\right)^2 
\left( \frac{A}{200}\right)
\left( \frac{2 \; 10^{-4}}\xi \right)^2
\left( \frac{\Delta M_t}{1\;{\rm MeV}}\right). 
\end{equation}

The above estimates give a rough idea on the number of 
collisions needed to reveal the $d^{\prime}$ peak. We see 
that light nuclei require smaller number of the collisions 
and therefore are better candidates for $d^{\prime}$ searches.
In order to obtain a more precise estimate and to check the 
$A$ dependence of the results, we perform in Sects. 4 and 5 Monte 
Carlo simulations for the $d^{\prime}$ production in central heavy 
ion collisions within the QMD  model. We analyse also additional cuts 
and show that they allow to reduce the value $N_C$ by a factor $1/2$.

\section{Simulation of $d^{\prime }$ production in heavy 
ion collisions}

The simulations were done within the framework of the 
Quantum Molecular Dynamics (QMD) 
model \cite{aich91}. QMD is a semi classical transport 
model which accounts for relevant 
quantum aspects like the Fermi motion of the nucleons, 
stochastic scattering processes 
including Pauli blocking in the final states, the creation and 
reabsorption of resonances
and the particle production. The time evolution of the 
individual nucleons is thereby governed 
by the classical equations of motion
\begin{eqnarray}
\frac{\partial {{\bf p}}_i}{\partial t} = 
- \frac{\partial H}{\partial {{\bf q}}_i},\quad 
\frac{\partial {\bf q}_i}{\partial t} =   
\frac{\partial H }{\partial {\bf p}_i}\quad  
\label{motion1}
\end{eqnarray}
with the classical $N-$body Hamiltonian $H$
\begin{equation}
H = \sum_i \sqrt{{\bf p}_{i}^2 + M_{i}^2} + 
\frac{1}{2} \sum_{i\neq j} \left( U_{ij} + U_{ij}^{\rm 
Yuk} + U_{ij}^{\rm Coul}\right).
\label{hamB}
\end{equation}
The Hamiltonian, Eq. (\ref{hamB}), contains two-body 
potential 
interactions which are finally determined as 
classical expectation values
from the local Skyrme forces $ U_{ij}$ supplemented by a 
phenomenological
momentum dependence,
\begin{equation}
\label{skyrme}
U_{ij} = \alpha\left(\frac{\rho_{ij}}{\rho_0}\right)
+\beta\left(\frac{\rho_{ij}}{\rho_0}\right)^{\gamma}
+\delta ln^2\left(\epsilon |{\bf p}_i -{\bf p}_j |^2 +1 \right)
\frac{\rho_{ij}}{\rho_0},
\end{equation}
and an effective 
Coulomb interaction $U_{ij}^{\rm Coul}$,
\begin{equation}
 \label{coul}
U_{ij}^{coul} =  {\left ( \frac{Z}{A}\right )}^2 
\frac{e^2}{|{\bf q}_i- {\bf q}_j |} 
erf\left(\frac{|{\bf q}_i-{\bf q}_j|}{\sqrt{4L}}\right).
\end{equation}
Here, $\rho_{ij}$ is a two-body interaction density defined 
as 
\begin{equation}
\rho_{ij} = \frac{1}{(4\pi L)^{\frac{3}{2}}}
e^{-( {\bf q}_{i}- {\bf q}_{j})^2 /4L}
\label{dens1}
\end{equation}
and $erf$ is the error function. The parameters $\alpha, 
\beta, \gamma, 
\delta, \epsilon$ of the Skyrme interaction, Eq. (\ref{skyrme}), 
are determined such as to
reproduce the saturation density 
($\rho_0 = 0.17$ fm$^{-3}$), the binding energy ($E_B=-
16$ MeV) for
normal nuclear matter at a fixed incompressibility 
$K_{\infty }$,
as well as the correct
momentum dependence of the real part of the 
nucleon-nucleus optical potential \cite{Ai86}. In the present 
calculations we use a soft equation of state 
with $K_{\infty }=$ 200 MeV corresponding to the 
momentum-dependent interaction (SMD), 
Eq. (\ref{skyrme}), which is 
known to reproduce reasonably well the reaction dynamics 
in the considered 
energy range. The Yukawa-type potential 
$ U_{ij}^{\rm Yuk}$ in Eq. (\ref{hamB}) serves to 
improve 
the surface properties and stability of the initialised nuclei. 

Two-body collisions are implemented by Monte Carlo 
methods,
the Pauli principle is thereby taken into account for the final 
states.
Nucleons while propagating collide with each other 
stochastically provided that the
distance between centroids of the two Gaussian wave 
packets 
is less than $d_{\rm min}=\sqrt {\sigma_{\rm tot}({\sqrt 
s})/\pi }$.
For the inelastic 
nucleon-nucleon channels, we include the relevant baryonic 
resonances, {\it  i.e.} $\Delta(1232)$ and 
$N^{*}(1440)$.
The pions originate from the respective resonance decays.  
In particular we include the resonance production and 
rescatterings 
in the inelastic $NN$ collisions: the one-pion decays of 
$\Delta$ and 
$N^{*}$, the two-pion decays of $N^{*}$, and the one-pion 
reabsorption processes. For details see Refs. 
\cite{fuchs97,uma98}. 

Since the production cross section of the $d^{\prime}$ 
dibaryons is small, these particles are 
treated perturbatively. The $d^\prime$ dibaryons are 
produced at kinematically favourable conditions in the 
$pp$ and $nn$ collisions with a probability 
$$
P = \frac{\sigma _{pp\rightarrow d^{\prime }\pi^+ 
}}{\sigma^{tot}_{NN}},
$$
while in the $pn$ collisions the $d^\prime$ dibaryons are 
produced with a probability $P/2$.
Here, $\sigma^{tot}_{NN}$ is the total (elastic + inelastic) 
$NN$ cross section and $\sigma 
_{pp\rightarrow d^{\prime }\pi^+ }$ is given by 
Eq. (\ref{sigmaff}). The results of our simulations for the
$d^{\prime}$ production rate in 1000, 50, 25, and 100 central 
collisions of the
ions $^{20}Ne$, $^{59}Ni$, $^{108}Ag$, and $^{197}Au$ 
are shown on Fig. 3 (a). The impact parameter $b$ is assumed
to be less, respectively, than $0.7$, $1.3$, $2$, and $3$ fm 
(minimal bias). In these intervals, the probability for the 
$d^{\prime}$ production is not sensitive to the values of $b$.
The $d^{\prime}$ production rate increases linearly with the 
atomic number $A$, and so our results can be presented in the form
\begin{equation}
\label{aaaa}
n_{d^{\prime}} \approx 10^{-4} \cdot A.
\end{equation}

\section{The invariant mass distribution of the background 
$pp\pi ^{-}$ events}

The number of the $d^\prime$ dibaryons
produced per single collision is much smaller 
than the number of the $NN\pi$ background events 
in the vicinity of the $d^{\prime}$ 
mass (according to Sect. 3). It is therefore reasonable to 
search for cuts which are able to enhance 
the $d^{\prime}$ signal. 

The first possible cut takes the fact into account that the 
$d^\prime$ is an isosinglet
state. Only $1/12$ of all $NN\pi$ events form an  
isosinglet state (we assume isotropy of 
the $NN\pi$ events in the isotopic space). The selection of 
isosinglet $NN\pi$ events could 
reduce the background by a factor of $12$. Such a selection 
is, however, impossible. Let us
select only $pp\pi^-$ events. The total reduction of the 
background is also $1/12$. Since the 
branching ratio of the $d^\prime\rightarrow pp\pi^-$ decay 
is $1/3$, the number of the 
$d^\prime$ events is reduced as well. The relative reduction 
of the number $N_C$ of collisions needed to reveal the $d^\prime$ peak is, 
according to Eq. (\ref {AAA}), $3^2/12 
= 3/4$. 

The second possible cut can be obtained by the selection of  
protons with an invariant mass 
$M_{pp} <  2m +  
15$ MeV. Such a cut not much influences the $d^\prime$ 
signal, since for the dominant part of 
the $d^\prime$ decays ($\approx 2/3$) the invariant mass of the two 
protons $M_{pp}$ is less than $2m + 15$ MeV. 
This is just the effect of the final-state interaction of 
two protons in the $d^\prime 
\rightarrow pp\pi^{-}$ decay. In Fig. 1, this effect is 
displayed in the fast increase of the 
function $\beta (T)$ at small nucleon kinetic energies $T$. 
The reduction of the background can be estimated as in 
Eq. (\ref{numberdensity1}) and gives
\begin{equation}
\label{yyyy}
\frac{T^{2}_{\pi}(M_{pp} = 2m) - T^{2}_{\pi}(M_{pp} = 2m 
+ 15 \; {\rm MeV})}
{T^{2}_{\pi}(M_{pp} = 2m)}{\;\approx \;}\frac{1}{3}
\end{equation}
where $T _{\pi}(M_{pp})$ is the kinetic energy of pions in 
the $NN\pi$ system for a given invariant mass 
 $M_{pp}$ of the two nucleons. In particular, $T 
_{\pi}(M_{pp} = 2m) = T _{\pi, 
\max}{\;\approx \;}45$ MeV. The relative reduction 
of the value $N_C$ is $(1/3) \cdot (3/2)^2 
= 3/4$.

The $d^\prime$ dibaryon decay products 
are distributed isotropically in the nucleus-nucleus center-of-mass 
frame, while the
nucleons in heavy ion collisions are produced with a higher 
probability  in 
the forward and backward directions. The angular 
distribution of the outgoing protons averaged over 100 
$^{197}Au\;+\;^{197}Au$ collisions is shown in Fig. 4. The cut in 
the polar angle of the beam can provide an additional 
reduction of the lower limit for the number of 
the collisions required to reveal the $d^{\prime}$ peak. The
background is reduced by a factor 2, while the number of the $d^{\prime}$
events is reduced by a factor $3/4$ (see below). The relative reduction 
of the number $N_C$ is small $(1/2) \cdot (4/3)^2 
= 8/9$. 

These three cuts when applied separately are not much efficient. 
Nevertheless, being combined they reduce the value $N_C$ by a factor 
$(3/4) \cdot (3/4) \cdot (8/9) = 1/2$.

The simulations of the $pp\pi^-$ background are  made 
within the framework of the QMD 
model for 1000, 50, 25, and 100 central 
collisions of the ions $^{20}Ne$, $^{59}Ni$, $^{108}Ag$, and $^{197}Au$ 
at the beam energy 
$1000$ MeV per nucleon. In order to exclude 
complex fragments from the analyses,
we take only protons into account 
which have no neighbouring nucleons closer than 3 fm 
in the final stage of the reaction. 
Here we trace the evolution of the reaction up to 
80 fm/c, a time where almost all resonances ($\Delta$ and $N^* (1440)$) 
have decayed into nucleons and pions. 

The invariant mass distribution of the $pp\pi^-$ events in 
the interval $2.02 < M_t < 3$ GeV
is shown in Fig. 5 (a) for one typical $^{197}Au\;+\;^{197}Au$ collision. 
This distribution has a maximum at 
$M_t \approx$ 2.3 GeV. 
The upper limit in the integral entering the denominator of 
Eq. (\ref{phalabel}) is therefore reasonable. 

In Fig. 5 (b) we show the invariant mass distribution of the 
events in the interval 2.02 - 2.1 
GeV, averaged over 100 collisions $^{197}Au\;+\;^{197}Au$. According to 
Eq. (\ref{numberdensity1}), the density of the $NN\pi$ 
events is proportional to the phase space of the $NN\pi$ 
system. Near the $NN\pi$ threshold, 
the phase space is proportional to square of the difference 
$M_t - M_{th}$
where $M_{th} = 2m + \mu$. We see that in this region 
the density of the $pp\pi^-$ 
events is fitted well 
by the curve $C (M_t - M_{th})^2$, in agreement 
with the 
expression (\ref{numberdensity1}). 
From the fit, we determine the coefficient 
$C = 2.6 \; 10^4$ 
GeV$^{-2}$. In vicinity of the $d^{\prime}$ dibaryon 
mass, the average number of the 
background $pp\pi^-$ events equals $\Delta N_t /N_C = 
55$. This number should be 
compared with the number in the right hand side of Eq. (\ref{III}). 
Here we got 1600 
$NN\pi$ events per 1 MeV or, equivalently, $1600/12 = 
130 \; pp\pi^-$ events 
per 1 MeV. The estimate of Sect. 3 is therefore in 
reasonable agreement with the results of 
the QMD simulations. 

In Fig. 5 (c) we show the distribution of the $pp\pi^-$ 
events for 100 
collisions $^{197}Au\;+\;^{197}Au$, grouped in 10 bins of 
a width of $\Delta M_t = 
1$ MeV around the $d^{\prime}$ 
dibaryon mass. The solid curve is the best linear fit to this 
distribution. This distribution is now further analysed in Fig. 6.

First, we construct the histogram (a): The number of the 
bins versus the deviation of the number 
of the $pp\pi^-$ events from the mean values determined 
by the solid straight line in Fig. 5 (c). 
In the vicinity of the $d^{\prime}$ mass, the number of 
$NN\pi$ triplets is about $\Delta N_t \approx 
5500$. From Fig. 6 (a) we deduce the RMS deviation 
$\sigma \approx 80$ in the number $\Delta N_t$. The 
Poisson law requires $\sigma \approx \sqrt {\Delta N_t} \approx 
75$. The background fluctuations are therefore distributed in 
agreement with the Poisson statistics.

In Fig. 6 (b), the same results are shown for the cut 
$M_{pp} <  2m + 15$ MeV. The total 
number of the $pp\pi^-$ events is reduced approximately 
by a factor 3: We obtain 
$\Delta N_t \approx 2000$ triplets, in the 
excellent agreement with the estimate of Eq. (\ref{yyyy}). The RMS  
deviation $\sigma \approx 65$ is somewhat 
larger. One can expect that the $\sigma$ is reduced by a 
factor $\sqrt 3 = 1.7$, while we get only a 
reduction by a factor of $\approx 1.3$. This deviation can, however, 
be explained by statistical fluctuations due to the relatively 
small number of the bins (= 10). 

The angular cut $|\cos\theta_p^{c.m.s.}| < 0.75$ reduces 
the background by a factor of two, while
the number of the detected $d^{\prime}$ events is reduced 
by a factor 3/4. The results 
for the double cut $M_{pp} <  2m + 15$ MeV and 
$|\cos\theta_p^{c.m.s.}| < 0.75$ are shown 
in Fig. 6 (c). Now we get only $\Delta N_t \approx 1100$ triplets. 
The RMS deviation $\sigma \approx 30$ 
is again in good agreement with the 
Poisson law, {\it i.e.} $\sigma \approx \sqrt {\Delta N_t} \approx 33$. 

To summarise, the selection of the $pp\pi^{-}$ events suppresses the 
background by a factor 
12. The second cut over the invariant masses of the two 
protons suppresses the background by a 
factor 3. The third cut over the proton scattering angles 
suppresses the background by a factor 
2. The number of the detected $d^{\prime}$ events 
decreases by a factor 1/3 in the 
first case, decreases by a factor 2/3 in the second case, and 
decreases by a factor 3/4 in the third case. 
The number $N_C$, which is the ratio between the 
background $\Delta n_t$ and square 
of the signal $n_{d^{\prime}}^2$ (see Eq. (\ref{AAA})), 
decreases by a factor $1/2$.

The fluctuations of the $pp\pi^{-}$ background summed 
up over all ion collisions are 
described reasonably well by 
the Poisson statistics, and so Eq. (\ref{AAA}), on which we base 
the estimate for the total number 
of collisions needed to reveal the $d^{\prime}$ 
peak, are apparently justified. 
 
At the same time, fluctuations of the $pp\pi^-$ background 
from one collision to 
another due to the variations of the impact parameter and 
due to variations of the coordinates and momenta of the 
nucleons inside of the colliding nuclei, are significant. These 
fluctuations are analysed in Fig. 7 with no cuts (a), with the 
cut 
$M_{pp} <  2m + 15$ MeV (b), and with the double cut 
$M_{pp} <  2m + 15$ MeV and 
$|\cos\theta_p^{c.m.s.}| < 0.75$ (c). The histograms (a) - 
(c) show the number of simulated collisions as a function of the number 
of all possible $pp\pi^{-}$ combinations per single collision 
in the bin at the $d^{\prime}$ dibaryon 
mass $M_{d^{\prime}}=2.063$ GeV with width $\Delta M_t = 1$ 
MeV. The reduction of the background by a factor 3 on the 
histogram (b) and further by a factor 2 on the histogram (c) 
as compared to the histograms (a) and (b), respectively, is 
clearly seen, while the RMS deviations are 2-3 times 
greater than one could expect from the Poisson distribution. 
It means that the distinction 
between different collisions has essentially a dynamical 
origin. If all triplets from all 
collisions are mixed, the Poisson distribution starts to 
work, as we discussed above.

With all the cuts taken into account, the number of the 
$pp\pi^{-}$ events in a bin $\Delta 
M_t = 1$ MeV around the $d^{\prime}$ mass per one 
collision equals $\Delta n_t =11$ (see Fig. 7 
(c)). This number should be compared with the number of 
the detected $d^{\prime}$ dibaryons in the channel $pp\pi^{-}$ 
$n_{d^{\prime}} \approx 0.02 \cdot (1/3) \cdot (2/3) \cdot (3/4) 
= 0.0033$. It means that the number of the $^{197}Au\;+\;^{197}Au$
collisions 
needed to reveal the $d^{\prime}$ 
peak at a $6 \sigma$ level is
\begin{equation}
\label{oooo}
N_C \geq n_{\sigma}^2 \frac{ \Delta 
n_t}{n_{d^\prime}^2}{\;\approx\;}3.6 \;10^7.
\end{equation}

The $A$ dependence of the number of the background events and 
of the ratio $\Delta n_t/n_{d^{\prime}}^2$ are analysed in 
Figs. 3 (b) and (c). The results of the QMD simulations
confirm essentially the naive estimates of Sect. 3: The value
$\Delta n_t$ increases as $A^3$, while the ratio 
$\Delta n_t/n_{d^{\prime}}^2$ increases linearly with $A$.
In the general case, Eq. (30) reads
\begin{equation}
\label{oooo}
N_C \geq 2 \;10^5 \cdot A.
\end{equation}

The limit on the value $N_C$ is inverse proportional to 
square of 
the $d^{\prime}$ production cross section and
depends linearly on the interval $\Delta M_t = \max 
(\Gamma_{d^{\prime}},(\Delta M_t)^{exp})$
where $(\Delta M_t)^{exp}$ is the resolution of the 
detectors. The $d^{\prime}$
cross section in Eq. (\ref{sigmaff}) 
is well defined provided that the 
signal seen at CELSIUS \cite {bro} 
is really connected to the $d^{\prime}$ dibaryon.  
Otherwise the used value of 
$\sigma_{pp \rightarrow d^{\prime} \pi^+}$ gives the upper 
limit for the $d^{\prime}$ production cross section, for the 
number $n_{d^{\prime}}$ of the detected 
$d^{\prime}$ dibaryons, and the minimum value for the 
number $N_C$.

\section{Summary}

We considered the production of the $d^\prime$ dibaryons 
in heavy ion
collisions. Due to smallness of the cross section for the 
$d^\prime$ dibaryon
production in the elementary $NN$ collisions, the 
production of the $d^\prime$
dibaryons in heavy ion collisions was considered as a 
perturbation to the
reaction dynamics. The number of the $d^\prime$ 
dibaryons was calculated using the 
QMD model for $^{20}Ne$, $^{59}Ni$, $^{108}Ag$, and $^{197}Au$
central collisions 
at the energy $1000$ MeV per 
nucleon. We obtained $n_{d^{\prime}} \approx 10^{-4} \cdot A$ dibaryons 
per single collision. This number was 
compared with the
background in the invariant mass distribution of the 
$pp\pi^-$ events. The
comparison showed that at least $2 \; 10^5 \cdot A$ 
collisions are needed to make the $d^{\prime}$ signal 
statistically significant. 
Light nuclei are more
promising candidates for the $d^{\prime}$ searches. 
If data on the $pp\pi^{-}$ invariant mass 
distribution from
earlier experiments are already available, the 
search for the $d^{\prime}$ peak can be made
both with light and heavy nuclei.
This is a realistic task to collect 
statistics sufficient to reveal 
the $d^{\prime}$ peak, and thus the 
search for the $d^\prime$ dibaryon seems to be a possibility 
which should be probed first by analysing the 
available heavy ion data accordingly.
\begin{acknowledgments}
Two of us (M.I.K and B.V.M.) are grateful to the Institute 
for Theoretical
Physics of University of T\"ubingen for kind hospitality. 
The work was supported 
by the Deutsche Forschungsgemeinschaft under the 
contract No. Fa67/20-1.
\end{acknowledgments}
\newpage
\section*{Appendix. The enhancement factor $\eta_F$ 
for reaction ${\normalsize d^{\prime}} \rightarrow NN\pi$ 
in the effective-range approximation}

The Jost function in the effective range approximation 
has the form (see [17], Ch.9)
$$
1/D(k) = \frac{k + i\alpha}{k + i\gamma}
$$
where 
$$
\frac{1}{2}r_e(\alpha - \gamma) = 1,\;\;\;\;\;\frac{1}{2}r_e\alpha\gamma = 1/a.
$$
Here, $a$ is the scattering length, $r_e$ fm the effective radius, and $k$ 
is the nucleon momentum in the center-of-mass frame of two nucleons. 
In our case,
$a = 23.7$ fm, $r_e = 2.67$ fm, and
$$
k = \sqrt {\frac{1}{2} M_{d^{\prime}}(T_{\pi,\max} - T_{\pi})}.
$$ 

In Fig. 1, the value $\beta(T) = \frac{1}{2}\ln{|D(0)/D(T)|^2}$ 
where $T = 2k^2/m$ is plotted (dashed curve).

The factor $\eta_F$ is obtained by averaging the value $|1/D(k)|^2$ over the
phase space of the $NN\pi$ system
$$
\eta_F = \frac {\int |1/D(k)|^2
\sqrt{T_{\pi}(T_{{\pi},{\max}} - T_{\pi})}dT_{\pi}}
{\int \sqrt{T_{\pi}(T_{{\pi},{\max}} - T_{\pi})}dT_{\pi}}.
$$
This expression can be transformed as follows
$$
\eta_F = \frac {4i}{\pi} \int_{C}^{} 
\frac{k_0^2z + \alpha^2}{k_0^2z + \gamma^2}
\sqrt{z(z - 1)}dz.
$$
The contour integral runs around the cut $[0,1]$, 
$Re(z)$ decreases when $Im(z)$ is positive and increases when 
$Im(z)$ is negative. 
Here,
$$
k_0 = \sqrt {\frac{1}{2} M_{d^{\prime}}T_{\pi,\max}}. 
$$
We move the contour $C$ to infinity and obtain
$$
\eta_F = 1 + 4(\alpha^2 - \gamma^2)(\sqrt{k_0^2 + \gamma^2} - \gamma)^2/k_0^4.
$$
Substituting $\alpha \approx 2/r_e >> \gamma \approx 1/a$ 
and $k_0 = 210$ MeV $>> \gamma$, we get 
$$
\eta_F \approx 1 + 16/(r_ek_0)^2 \approx 3.
$$
For $T=800$ MeV and $k=\sqrt{\frac{1}{2}mT} = 600$ MeV, 
$\eta_I = |1/D(k)|^2 \approx 1$.

\newpage
\section*{Figure Captions}

\begin{itemize}
\item[Fig.\,1:] The function $\beta (T) =\frac 12\ln \left( \eta 
(T)/\eta (0) \right)$ versus the 
laboratory kinetic energy $T$ of the incoming nucleon. The solid curve
represents the calculation with the Jost function (7), while the dashed
curve is due to the effective-range approximation described in Appendix.

\item[Fig.\,2:] The cross section $\sigma _{pp\rightarrow 
d^{\prime}\pi ^+}$ versus the 
laboratory kinetic energy $T$ from the $d^{\prime}\pi$ 
threshold up to $1000$ MeV for the reaction 
$pp\rightarrow d^{\prime}\pi^+$.  

\item[Fig.\,3:] (a) The $d^{\prime}$ production rate $n_{d^{\prime}}$ 
versus the atomic number $A$
of the colliding nuclei $^{20}Ne$, $^{59}Ni$, $^{108}Ag$, and $^{197}Au$
at 1 A$\cdot$GeV incident energy. The
errors display variations of the value $n_{d^{\prime}}$ due to variations 
of the impact parameter and initial coordinates and momenta of nucleons 
in the colliding
nuclei. The solid straight line is the linear fit of the 
value $n_{d^{\prime}}$.
(b) The number $\Delta n_t$ of the $pp\pi^{-}$ triplets per 1 MeV of 
the $pp\pi^{-}$ invariant mass
in vicinity of the $d^{\prime}$ peak versus atomic number $A$
of the colliding nuclei. The errors
give statistical fluctuations. The solid curve is the $A^3$ fit
of the value $\Delta n_t$.
(c) The ratio between the background $pp\pi^{-}$ events and square of 
the number of the $d^{\prime}$ 
dibaryons versus atomic number $A$. The 
solid straight line is the linear fit of the ratio 
$\Delta n_t/n_{d^{\prime}}^2$.

\item[Fig.\,4:] The proton polar angular distribution in the 
center-of-mass system of the 
colliding nuclei in central ($b< 3$ fm) $^{197}Au\;+\;^{197}Au$ 
collisions at 1 A$\cdot$GeV incident energy.

\item[Fig.\,5:] The distribution of the $pp\pi^-$ background 
events in central ($b< 3$ fm) $^{197}Au\;+\;^{197}Au$ collisions at 
1 $A\cdot$GeV incident energy.
(a) The $pp\pi^-$ background in the interval $2 \div 3$ 
GeV of the invariant mass $M_t$ for a single 
typical collision. (b) The $pp\pi^-$ background near threshold 
averaged over $N_C = 100$ collisions. 
According to Eq. (\protect\ref{numberdensity1}), 
the distribution has a simple threshold behaviour: 
$\Delta N_t = C (M_t - 
M_{th})^2$. It is seen that the parabolic curve is an 
excellent fit to the distribution near 
the threshold. (c) The total 
$pp\pi^-$ background at threshold (see scale of $M_t$) 
for $N_C = 100$ collisions. The straight solid line gives the best fit for 
these 10 bins. The deviations of the 
$\Delta N_t$ from the mean values indicated by the solid 
line are in good agreement with the 
Poisson distribution (see Figs. 6 and discussion in the 
text). The value $\Delta N_t = ( dN_t 
/dM_t ) \Delta M_t$ is the number of the $pp\pi^-$ events 
in the interval $\Delta 
M_t$. In the histograms (a), (b), and (c), $\Delta M_t =$ 
10, 1, and 1 MeV, respectively.

\item[Fig.\,6:] The numbers of the bins of Fig. 5 (c) with the 
fixed deviations from the mean 
values of the $\Delta N_t $, indicated by the solid line, with 
no cuts (a), with the 
cut over the two-proton invariant masses $M_{pp}< 
2m+15$ MeV (b), and with the double cut 
over the two-proton invariant masses $M_{pp}< 2m+15$ 
MeV and the proton scattering angles 
$|\cos\theta_p^{c.m.s.}| < 0.75$ in the center-of-mass 
system of the colliding nuclei (c). The 
RMS deviations $\sigma$ = 80, 65, and 30 should be 
compared with the 
mean values $\Delta N_t $ = 5500, 2000, and 1100  in the 
cases (a), (b), and (c), respectively 
(the mean values $\Delta N_t/N_C $ with $N_C = 100$
can be read off 
from the upper right corners of Figs. 7 
(a) - (c)). The fluctuations of the background $\sigma 
\approx \sqrt{\Delta N_t}$ 
do not contradict to the Poisson statistics.

\item[Fig.\,7:] The distribution of 100 simulations 
of $^{197}Au\;+\;^{197}Au$ collisions as a function of 
the number of the $pp\pi^-$ combinations (triplets) per 
single collision in the region $M_{d^{\prime}}\pm 
0.5$ MeV with no cuts (a), with 
the cut over the two-proton invariant mass $M_{pp}< 
2m+15$ MeV (b), and with the double 
cut over the two-proton invariant mass $M_{pp}< 
2m+15$ MeV and the proton scattering 
angles $|\cos\theta_p^{c.m.s.}| < 0.75$ in the center-of-
mass system of the colliding nuclei (c). 
\end{itemize}
\newpage
\begin{figure}[h]
\begin{center}
\leavevmode
\epsfxsize = 15cm
\epsffile[85 400 460 690]{fig1.eps} 
\end{center}
\caption{\label{fig1}
}
\end{figure}
\begin{figure}[h]
\begin{center}
\leavevmode
\epsfxsize = 15cm
\epsffile[85 400 460 690]{fig2.eps}
\end{center}
\caption{\label{fig2}
}
\end{figure}
\begin{figure}[h]
\begin{center}
\leavevmode
\epsfxsize = 13cm
\epsffile[90 60 505 752]{fig16.eps}
\end{center}
\caption{\label{fig3}
}
\end{figure}
\begin{figure}[h]
\begin{center}
\leavevmode
\epsfxsize = 13cm
\epsffile[130 340 460 670]{fig4.ps}
\end{center}
\caption{\label{fig4}
}
\end{figure}
\begin{figure}[h]
\begin{center}
\leavevmode
\epsfxsize = 13cm
\epsffile[110 140 460 690]{fig5.ps}
\end{center}
\caption{\label{fig5}
}
\end{figure}
\begin{figure}[h]
\begin{center}
\leavevmode
\epsfxsize = 13cm
\epsffile[110 140 460 690]{fig6.ps}
\end{center}
\caption{\label{fig6}
}
\end{figure}
\begin{figure}[h]
\begin{center}
\leavevmode
\epsfxsize = 13cm
\epsffile[110 140 460 690]{fig7.ps}
\end{center}
\caption{\label{fig7}
}
\end{figure}

\end{document}